\documentclass[twocolumn]{aastex701}
\usepackage{amsmath}
\usepackage{amssymb}
\usepackage{mathtools}
\usepackage{bm}
\usepackage{tabularx}
\usepackage{booktabs}

\begin{document}

\title{1D PIC Simulations of Resonant Scattering–Driven Pair Cascades in Magnetar Magnetospheres}

\author[orcid=0000-0002-5349-7116]{Jens F. Mahlmann}
\affiliation{Department of Physics \& Astronomy, Wilder Laboratory, Dartmouth College, Hanover, NH 03755, USA}
\email[show]{jens.f.mahlmann@dartmouth.edu} 
\author[0000-0001-7801-0362]{Alexander A. Philippov}
\affiliation{Department of Physics, University of Maryland, College Park, MD 20742, USA}
\affiliation{Physics Department, Stanford University, Stanford, CA 94305, USA}
\email[]{email} 
\author[0000-0002-8599-8847]{Adrien Soudais}
\affiliation{Department of Physics \& Astronomy, Wilder Laboratory, Dartmouth College, Hanover, NH 03755, USA}
\email[]{email} 
\author[orcid=0000-0001-5660-3175
]{Andrei M. Beloborodov}%
\affiliation{Department of Physics, Pupin Hall, Columbia University, New York, NY 10027, USA}
\affiliation{Max-Planck Institute for Astrophysics, Garching, 85741, Germany}
\email[]{email} 
\author[orcid=0000-0002-1227-2754
]{Lorenzo Sironi}
\affiliation{Department of Astronomy and Columbia Astrophysics Laboratory, Columbia University, New York, NY 10027, USA}
\affiliation{Center for Computational Astrophysics, Flatiron Institute, New York, NY 10010, USA}
\email[]{email} 

\correspondingauthor{Jens F. Mahlmann}

\begin{abstract}
Hard X-rays in persistent magnetar emission originate from radiative processes in pair-loaded magnetospheric plasma, yet self-consistent kinetic simulations of such radiation-rich systems remain limited. We perform 1D particle-in-cell simulations of pair creation mediated by resonant inverse Compton scattering (RICS) along isolated field lines in twisted magnetospheres, capturing particle acceleration and radiative drag from first principles. The drag force, which strongly depends on the particle Lorentz factor and location, plays a key role in the magnetospheric circuit. Strong RICS drag impedes the flow of electrons and positrons, particularly along extended field lines. Nevertheless, the plasma sustains the circuit by self-organizing into a single accelerating gap near the star in one hemisphere. The accelerated plasma flow from the gap becomes loaded with copious electron–positron pairs in extended RICS zones, and also carries ions extracted from the star and accelerated in the gap. When radiative drag stops the pair-loaded flow at the magnetic equator, the ion component transfers momentum through streaming instabilities, delivering a small lepton population into the opposite hemisphere and sustaining the circuit. The gap confined to a single (anode) hemisphere implies asymmetric hard X-ray production, possibly detectable in phase-resolved magnetar spectra.
\end{abstract}

\keywords{Magnetars (992); Neutron stars (1108); Plasma astrophysics (1261); X-ray sources (1822); High energy astrophysics (739)}


\section{Introduction}

Magnetars are young, highly magnetized neutron stars (NSs) generating persistent and transient X-ray emission \citep{Kaspi2017, Esposito2020}. Their inferred magnetic fields exceed the characteristic quantum electrodynamics field $B_{\rm QED}=4.4\times 10^{13}\,{\rm G}$. Dissipation of magnetic energy is believed to power the observed X-rays, including long-lived emission with luminosities of $10^{34}-10^{36}{\,\rm erg\,s^{-1}}$. The spectrum of this persistent (long-lived) emission typically has a quasi-thermal component near $1\,{\rm keV}$, and a nonthermal tail extending beyond $100\,{\rm keV}$. While thermal emission may come from the magnetar surface, nonthermal X-rays must be emitted by the surrounding magnetosphere. An expected emission process is resonant inverse Compton scattering \citep[RICS, e.g.,][]{Thompson2002,Beloborodov2013,Wadiasingh2018,Harding2025}. RICS emission can be calculated for a given pattern of plasma motion in the twisted magnetosphere. Finding this self-organized pattern is a key challenge, which requires understanding of how and where electrons and positrons, $e^\pm$, are created. \cite{Beloborodov2013} proposed a model of $e^\pm$ outflow from an electric gap near the neutron star and decelerated by RICS. The model was found to fit the phase-resolved spectra of observed nonthermal X-rays \citep{Hascooet2014}.

A self-organized plasma state in a current-carrying (twisted) magnetosphere, persisting over observed timescales of months to years, involves a delicate balance in the plasma circuit \citep{Beloborodov2007,Beloborodov2013_eflows}. Charge carriers of density $n$ move with velocity $v_\parallel$ along magnetic field lines to sustain a magnetospheric current density $j\sim nev_\parallel$, where $e$ is the elementary charge. Some \emph{primary} particles can be supplied by the atmospheric layer on the stellar surface. However, most of magnetospheric plasma is provided by $e^\pm$ creation, triggered by particle acceleration in regions of unscreened electric fields, which must exist somewhere in the circuit. Particles also experience significant radiative losses, which regulate $v_\parallel$. \citet{Beloborodov2011} predicted two distinct zones set by RICS. In the inner corona, where the magnetic field satisfies $B \gg B_{\rm QED}$,
pair creation is efficient, while RICS only mildly decelerates the flow. The pair plasma therefore maintains a moderate multiplicity, $\mathcal{M} \approx nec/j$, where $c$ is the speed of light. In the outer corona, where $B \lesssim B_{\rm QED}$, the flow becomes drag-force-dominated and decelerates strongly when approaching the magnetic equator. The $e^\pm$ outflow along such extended magnetic field lines is expected to have a high multiplicity as a result of the RICS-mediated pair-creation avalanche.

Kinetic particle-in-cell (PIC) simulations have become a powerful tool to generate insight into relativistic plasmas around compact objects. Despite the substantial challenges of resolving the relevant kinetic and radiative scales, they have been successfully used to model particle acceleration, plasma supply, and coherent emission in pulsar magnetospheres \citep[][and references therein]{Philippov2022}. First-principles models of global plasma kinetics in magnetar magnetospheres are challenging due to the separation between radiative-reaction and plasma scales. The first kinetic model of an $e^\pm$ discharge in magnetars used a 1D setup \citep{Beloborodov2007}, and later \citet{Chen2017} simulated pair production in 2D (axisymmetric) twisted magnetospheres.

This Letter first summarizes key challenges of PIC simulations of the magnetar circuit, including limitations on RICS drag arising from the required resolution of plasma scales. Then, we simulate the magnetosphere as a collection of 1D (curved) field lines with a dipolar geometry, and find the self-organized plasma state on each field line. The paper is organized as follows. Section~\ref{sec:plasma_circuit} reviews how magnetospheric currents accelerate plasmas along field lines, the basics of RICS (\ref{sec:rics_basics}), and the scales induced by the radiative drag (\ref{sec:rics_scales}). We describe our simulations in Section~\ref{sec:simulations} and present their results in Section~\ref{sec:results}, focusing on the location of pair production (\ref{sec:phase_structure_gaps}), as well as the resulting pair multiplicity distribution and its scaling with the ion-to-electron mass ratio (\ref{sec:mass_ratio_multiplicity}). We discuss our results in Section~\ref{sec:discussion}. The Appendix reviews RICS energetics (Appendix~\ref{app:rics_cross_section}), specifics of our simulation setup (Appendix~\ref{app:setup}), and the dependence of the circuit on the RICS drag-time limit (Appendix~\ref{app:qedprocesses}).

\section{The Magnetar Plasma Circuit}
\label{sec:plasma_circuit}

Displacements of the solid magnetar surface can shear magnetic field lines \citep{Thompson_Duncan_1995MNRAS.275..255}. The toroidal shear of axisymmetric dipole fields induces a twist $\psi$, measuring the difference $\psi=\phi_1-\phi_0$ of a field line's footpoint positions in spherical coordinates $(r,\theta,\phi)$. The current required to sustain such a magnetospheric twist is approximately
\begin{align}
 j_\psi=\frac{c}{4\pi}|\nabla\times\mathbf{B}|\approx\frac{c}{4\pi}\frac{B_\ast}{R_\ast}\psi\,.
 \label{eq:targetcurrent}
\end{align}
Here, $R_\ast$ denotes the stellar radius, and $B_\ast$ is the surface magnetic field. The target current $j_\psi$ requires a charge-separated flow of density $n_\psi=j_\psi/(ec)$. The energy available to leptons before reacting with the radiation field is limited by the double-layer potential, $V_{\psi}$ \citep[][]{Carlqvist1982,Beloborodov2007}:
\begin{align}
\begin{split}
 \gamma_{\psi}&=\frac{eV_{\psi}}{m_e c^2}= \kappa \frac{R_\ast}{d_\psi} \,,
 \label{eq:gamma_circuit}
\end{split}
\end{align}
where $d_\psi^2 = m_e c^2/(4\pi n_\psi e^2)$ is the cold plasma skin-depth associated with the circuit, $m_e$ and $m_i$ are
the electron and ion masses, respectively, and $\kappa=[1+(m_i/m_e)^{1/2}]/2$ for electron--ion plasmas. We assumed an acceleration length $L_{\rm acc}\approx R_\ast$ to account for transverse gradients in the current bundle that limit the accelerating potential. Ohmic dissipation, $\mathbf{E}\cdot\mathbf{j}=E_\parallel j_\psi$, becomes substantial when the potential of the plasma-filled circuit, $V_{\rm c}\equiv -\int E_\parallel\,{\rm d}l$, reaches a non-negligible fraction of the double-layer potential, $V_\psi$. Here, $E_\parallel$ is the electric field along the magnetic field, and the integral is taken along the field line with differential arc length ${\rm d}l$. Strong $E_\parallel$ dissipate the magnetospheric twist by transferring magnetic energy to particles, resulting in decreasing currents $j_\psi$ and suppressed RICS \citep[][section 2.2]{Beloborodov2007}. Observations of persistent magnetar emission imply slow dissipation timescales \citep[e.g.,][]{Esposito2020,Younes2022,Younes2025}. We therefore define `ignited' magnetar circuits as those that sustain the plasma flows required to maintain the magnetospheric twist in the presence of radiation reaction without strong dissipation. This Letter identifies steady-state solutions of this kind.

\subsection{Resonant Inverse Compton Scattering (RICS)}
\label{sec:rics_basics}

In this section, we review the resonance condition for RICS and its implications for the energy of upscattered photons. Leptons interact with thermal photons emitted from the magnetar surface. In strong magnetic fields, radiative feedback is mediated by RICS through the first excited Landau state, with rest-frame energy
\begin{align}
 \frac{E_B}{m_e c^2}=\left(1+\frac{2B}{B_{\rm QED}}\right)^{1/2}\label{eq:elandau}\,,
\end{align}
where $B_{\rm QED} = m_e^2 c^3 / \hbar e$. A photon of energy $\hbar\tilde{\omega}_X$ excites a lepton to $E_B$ if it matches the resonance condition
\begin{align}
 \tilde{\omega}_X=\omega_B=\frac{eB}{m_e c}\,.\label{eq:resonancerest}
\end{align}
Here, tilde denotes quantities in the rest frame of a lepton moving along the magnetic field with Lorentz factor $\gamma=(1-\beta^2)^{-1/2}$. For a photon with an angle $\mu=\cos\theta$ to the lepton trajectory, this corresponds to
\begin{align}
 \omega_X=\frac{\tilde{\omega}_X}{\gamma(1-\beta\mu)}=\frac{B/B_{\rm QED}}{\gamma(1-\beta\mu)}\frac{m_e c^2}{\hbar}\,,\label{eq:resomega}
\end{align}
where $\hbar\omega_B=(B/B_{\rm QED})m_e c^2$. The resonance condition for lepton excitation to the first Landau level is
\begin{align}
 \gamma_{\rm res}= 511\times\frac{B/B_{\rm QED}}{(1-\beta\mu)}\left(\frac{1\,{\rm keV}}{E_X}\right)\,,
 \label{eq:resgamma}
\end{align}
using the photon energy $E_X$ in the NS frame. RICS has an excitation phase, where leptons occupy the first Landau level, and a de-excitation phase, where an energetic photon is emitted. The energy of the up-scattered photons, $E_\gamma$, depends on the emission angle (see Appendix~\ref{app:energy_upscattered}) and can be estimated in the NS frame:
\begin{align}
 \gamma_{\rm res}\left(1-\frac{m_e c^2}{E_B}\right)<\frac{E_\gamma}{m_e c^2}< \gamma_{\rm res}\left(1-\frac{m_e^2 c^4}{E_B^2}\right)\,.
\end{align}
Using Equation~(\ref{eq:resgamma}), one obtains
a constraint on the magnetic field strength required to produce photons with $E_\gamma\geq 2m_ec^2$, sufficient for pair production:
\begin{align}
 b\equiv\frac{B}{B_{\rm QED}}\gtrsim 0.063\times\left(\frac{E_X}{1\,{\rm keV}}\right)\,.
 \label{eq:blimit}
\end{align}
Here, we assume $b\ll 1$ and expand $1-m_ec^2/E_B\approx b$.

\subsection{Scales of the RICS Drag Force}
\label{sec:rics_scales}

In this section, we derive the RICS-induced drag force and associated RICS drag timescale considering only the direct thermal photon field from the stellar surface and neglecting re-processed photons scattered by the magnetospheric plasma. We then compare the RICS drag length to the plasma skin-depth of the magnetosphere. The RICS rate for a thermal background of temperature $T$ follows from Planck's law (here for perpendicular photon polarization, with $I_\parallel=0$):
\begin{align}
 I_\omega=\frac{\hbar\omega^3}{8\pi^3c^2}\left[\exp\left(\frac{\hbar\omega}{k_BT}\right)-1\right]^{-1}\,.
\end{align}
The spectral energy density is $u_\omega = I_\omega/c$. Using $\lambdabar=\hbar/m_e c$ to calculate the photon number density $n_\omega$ in a specific angular frequency interval, the number of scattering events per unit time and lepton is
\begin{align}
\begin{split}
 \dot{N}_{\rm sc}&=c \int\text{d}\Omega\int\text{d}\omega\,n_\omega{\sigma}_{\rm res}\\
 &=\frac{r_e}{4\pi\lambdabar^2}\frac{c}{\gamma}\int\text{d}\Omega \left[\frac{\hbar\omega_{X}}{m_e c^2}\right]^2\frac{1}{\exp\left[{\frac{\hbar\omega_{X}}{k_B T}}\right]-1}\,,
 \label{eq:scatteringNdot}
\end{split}
\end{align}
where ${\sigma}_{\rm res}$ is the RICS cross section \citep[see][]{Beloborodov2013}. We evaluate the radiative drag force $\mathcal{F}\equiv \dot{N}_{\rm sc}\Delta P$ for the de-excitation momentum loss of leptons, $\Delta P=\gamma\tilde{\mu}\hbar\omega_{B}/c$ (lab frame), and $\tilde{\mu}=(\mu-\beta)/(1-\beta\mu)$:
\begin{align}
 \mathcal{F}=\frac{r_e \Theta^3}{4x^2\lambdabar^2}\gamma \left(\mu-\beta\right)g(y)m_e c^2\,.\label{eq:contdrag}
\end{align}
Here, we use $x=r/R_\ast$, $\Theta=k_BT/(m_e c^2)$, $g(y)=y^3/\left(\exp\left[y\right]-1\right)$, $y=\hbar\omega_{X}/(k_BT)$, and a dimensionless drag coefficient \citep[][equation~B8]{Beloborodov2013}:
\begin{align}
 \mathcal{D}=\frac{r}{c}\frac{1}{p}\frac{{\rm d}p}{{\rm d}t}=\frac{r}{c}\frac{1}{p}\mathcal{F}\,.
 \label{eq:dcoeff}
\end{align}
The RICS drag timescale is $\tau_{\rm RICS}=p/|{\rm d}p/{\rm d}t|$, and substitution into Equation~(\ref{eq:dcoeff}) yields $\mathcal{D}=(r/c)\tau_{\rm RICS}^{-1}$. We write the RICS drag length as $l_{\rm RICS}=c\tau_{\rm RICS}=r/\mathcal{D}$. $\mathcal{D}$ measures the fractional momentum loss over a light crossing time $r/c$. The radiative drag pushes a particle towards a velocity attractor $\bar{\beta}=\mu$, with $\bar\gamma=1$ when the flow reaches the equator plane. For small deviations from the attractor momentum $\tilde{p}=\bar{\gamma}\bar{\beta} m_ec$, linearization around the equilibrium in Equation~(\ref{eq:contdrag}) implies
\begin{align}
\label{eq:D}
 \mathcal{D}=\frac{\alpha}{4x}\frac{R_\ast}{r_e}\Theta^3 \frac{g(\tilde{y})}{\bar{\gamma}^2}\left(1-\frac{p}{\tilde{p}}\right)\equiv\bar{\mathcal{D}}\left(1-\frac{p}{\tilde{p}}\right)\,.
\end{align}
For typical values of $\bar{\mathcal{D}}$ in the magnetospheric equator with $R_\ast=10^6\,{\rm cm}$ and maximum radiation density at $g(\tilde{y})\approx 1.4$ we estimate the RICS drag length
\begin{align}
 l_{\rm RICS}=\frac{r}{\bar{\mathcal{D}}}\approx 200\times\bar{\gamma}^2\left(\frac{x}{10}\right)^{2}\left(\frac{1\,{\rm keV}}{k_BT}\right)^{3}{\rm cm}\,.
\end{align}
We can compare this length scale to the approximate plasma skin-depth. The plasma density roughly scales with the magnetic field strength, $n\propto B\propto r^{-3}$. For typical densities $n\sim10^{37}r^{-3}$ \citep[see][]{Beloborodov2013,Beloborodov2020}, we find:
\begin{align}
 d_{\rm p}=\left(\frac{m_e c^2}{4\pi n_0 e^2}\right)^{1/2}\approx 5.4\times 10^{-3}\left(\frac{x}{10}\right)^{3/2}{\rm cm}\,.
\end{align}
At the magnetospheric equator, the scale hierarchy is
\begin{align}
d_{\rm p}\ll l_{\rm RICS}= c\tau_{\rm RICS}\sim 1\,{\rm m}
\ll r\sim 100\,{\rm km}\,,
\end{align}
for the fiducial parameters and $\bar{\gamma}\approx 1$. The dimensionless parameter of key importance for the magnetospheric circuit is $\eta\equiv \tau_{\rm RICS}\,\omega_p=l_{\rm RICS}/d_{\rm p}$, where
\begin{align}
 \eta\approx 3.8\times 10^4\left(\frac{x}{10}\right)^{1/2}\left(\frac{1\,{\rm keV}}{k_BT}\right)^{3}\,.
 \label{eq:tcool}
\end{align}
In practice, numerical simulations do not reach the huge ratios $l_{\rm RICS}/d_{\rm p}$ and $r/l_{\rm RICS}$ found in real magnetar magnetospheres. However, limiting $\eta>\eta_{\rm min}\approx 10^3$ in our models appears sufficient to achieve the correct physical regime of the circuit, as discussed below.

\section{Simulations}
\label{sec:simulations}

\begin{figure*}
 \includegraphics[width=0.99\linewidth]{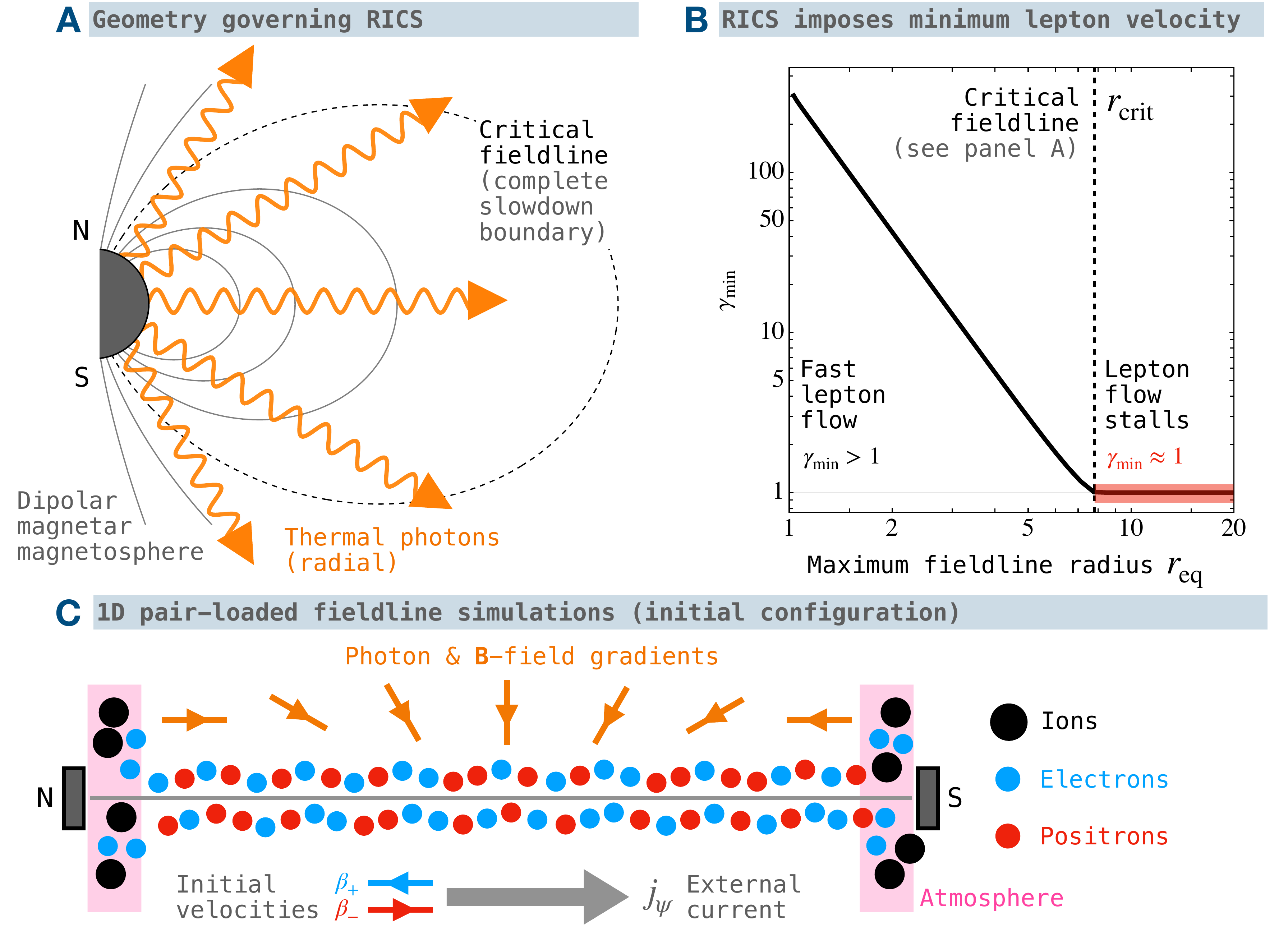}
 \caption{Schematic illustration of the radiation-rich dipolar magnetar magnetosphere and simulation setup for $k_{\rm B}T=1\,{\rm keV}$ and $B_\ast/B_{\rm QED}=10$. Panel A shows the dipole magnetosphere, with thermal photons emerging radially from the stellar surface (orange arrows). Panel B shows the minimum velocity of the lepton flow obtained for RICS of test particles, with plasma coming to a complete slow-down beyond the critical radius (also indicated in Panel A, and see Appendix~\ref{app:rics_cross_section}). Panel C shows the initial configuration of the 1D toy model of field lines in an effectively dipolar geometry (Section~\ref{sec:simulations}). A gravitationally supported atmosphere of ions and electrons is maintained at the domain boundaries (i.e., the stellar surface). The domain is initially filled with electrons and positrons sustaining an external current $j_\psi$. The RICS reaction is modeled by prescribing gradients of photon properties and magnetic field strength consistent with the dipolar geometry (Appendix~\ref{app:1D_geometry}).}
 \label{fig:FIGURE1}
\end{figure*}

We use \textsc{Tristan-MP.v2} \citep{tristanv2} to conduct 1D PIC simulations of relativistic plasma flows along dipolar magnetic field lines with RICS drag induced by photons emitted from a magnetar. The 1D domain resolves field lines of length $L$ with $N_x$ cells and a fixed resolution of $\Delta l/R_*=\left(L/R_\ast\right)N_x^{-1}=6.67\times10^{-5}$ (see Figure~\ref{fig:FIGURE1}, panel C). Each update advances the simulation by a timestep $\Delta t=\left(c_0/c\right)\Delta l$, where $c_0/c=0.5$ is an effective CFL factor. The fiducial magnetization at density $n_0$ is $\sigma_0=1.6\times 10^{8}$, with 48 cells per skin-depth. Normalized to code-units, the density scales as
\begin{align}
 \frac{n_\psi}{n_0}=\frac{j_\psi}{n_0ec}= 
 \left(\frac{d_0}{R_\ast}\right)\sigma_0^{1/2}\psi\,,
\end{align}
and the acceleration potential can be written as
\begin{align}
 \gamma_{\psi}&= \kappa \frac{R_\ast}{d_0}\left(\frac{j_\psi}{n_0ec}\right)^{1/2}\hspace{-4pt} = \kappa\left(\frac{R_\ast}{d_0}\sigma_0^{1/2}\psi\right)^{1/2}\,.
\end{align}
Particle and field boundary conditions at the domain ends are treated separately. We impose atmospheric conditions for particles: electrons and ions of mass ratio $m_i/m_e$ and temperature $T/m_e c^2=0.01$ are continuously injected up to a local plasma density threshold prescribed by gravitationally supported hydrostatic equilibrium with up to $n/n_\psi \approx 18$ close to the star. The exponentially decaying atmospheric density has a scale height of a few (local) skin-depths. Resolving the skin-depth puts limits on the maximum atmospheric density at the available resolution, but choosing $n/n_\psi \gg 1$ ensures that enough atmospheric particles are available and that boundary parameters (e.g., temperature of newly injected particles) do not affect the magnetospheric equilibrium. The atmosphere's `puffy' high-density layers serve as a source of plasma for the domain interior (Figure~\ref{fig:FIGURE1}, and Appendix~\ref{app:atmo_boundary}). EM fields have perfect conductor boundaries.

We prescribe a static external current $\mathbf{j}_{\rm ext}=j_\psi\mathbf{\hat{x}}$ in the entire domain, using $j_\psi/(n_0ec)=28$, corresponding to a twist $\psi\approx 0.7$. This external current is applied by solving ${\rm d}\mathbf{E}/{\rm d}t = 4\pi\left(\mathbf{j}_{\rm ext} - \mathbf{j}\right)$, effectively freezing the magnetospheric twist and allowing the plasma current $\mathbf{j}$ to self-consistently respond to it \citep{Beloborodov2007,Timokhin2013}. Initially, at $t=0$, a homogeneous electron-positron ($+/-$) pair plasma streaming with velocity $\beta_\pm=\pm c/\mathcal{M}$ fills the domain, where $\mathcal{M}$ denotes the pair multiplicity and we use $\mathcal{M}\approx nec/j_\psi$. We model atmospheric electrons and ions separately from secondary electron–positron pairs, with leptons undergoing RICS interactions. Throughout the domain, pairs are annihilated once the total density approaches the skin-depth limit $n_{\rm max}$, ensuring that the local plasma skin-depth is resolved by at least one grid cell. This numerical constraint also reflects the physical expectation that annihilation becomes more efficient where the flow decelerates and densities increase. We track the plasma response to realistic RICS drag, combined with a simplified treatment of pair production. In this toy model, the derivation of the radiative drag on each particle uses the geometry, magnetic field strength, and photon distribution of a dipole magnetosphere (Appendix~\ref{app:1D_geometry}) with a surface field strength $B_\ast/B_{\rm QED}=10$. The radiation kernel evolves momenta $\mathbf{\hat{p}}=\gamma\boldsymbol{\beta}$ with a normalized drag-force $\hat{\mathcal{F}}=\mathcal{F}\times R_\ast/(m_e c^2)$, accounting for the rescaling of spatial coordinates to units of $R_\ast$.

Our numerical scheme distinguishes two RICS responses. For $b> 0.09$, up-scattered energetic photons can pair produce (Equation~\ref{eq:blimit}). We define an approximate scattering probability $p_{\rm sc}=\dot{N}_{\rm sc}\,\Delta t$ for every particle. When activated, RICS events induce lepton recoil via excitation and de-excitation of the first Landau level (Appendix~\ref{app:rics_cross_section}). When the energy of an up-scattered photon exceeds the pair production threshold, $E_\gamma \ge 2 m_e c^2$, we immediately inject an electron-positron pair, sharing the photon momentum equally. Pair production is limited by $n_{\rm max}$ to preserve the skin-depth constraint.

For $b<0.1$, photons escape without appreciable pair production and need not be retained on the grid. The total force on particles is the sum of the Lorentz force and the radiative drag force. We use Strang splitting to apply the continuous radiative drag force $\hat{\mathcal{F}}$: a drag update over $\Delta t/2$, the particle pusher's Lorentz-force update over $\Delta t$, and a second drag update over $\Delta t/2$. We limit the applied drag force by
\begin{align}
 \left|\frac{\Delta p}{p_0}\right|\approx\left|\frac{\hat{\mathcal{F}}\times \Delta t}{2p_0}\right|\lesssim\frac{\Delta t/2}{\tau_{\rm min}}\,,
\end{align}
where we choose $\eta_{\rm min}\equiv \tau_{\rm min}\omega_{\rm p}= 10^3\times(n_\psi/n_0)^{1/2}\gg 1$ (see Equation~\ref{eq:tcool}) and $\Delta p=p_1-p_0$ is the difference to the initial momentum $p_0$. When $|\Delta p/p_0|<\Delta t/(2\tau_{\rm min})$, we calculate $p_1$ by solving the implicit midpoint equation
\begin{align}
 p_1=p_0+\frac{\Delta t}{2}\times \frac{\hat{\mathcal{F}}(p_0)+\hat{\mathcal{F}}(p_1)}{2}\,,\label{eq:implicit}
\end{align}
with an appropriate Newton-Raphson scheme. Otherwise, we drive the momentum to its target value $\tilde{p}=\mu/(1-\mu^2)^{1/2}$ by applying the following damping:
\begin{align}
 p_1 &=\tilde{p}+(p_0-\tilde{p})\times\exp\left[-\frac{\Delta t}{2\tau_{\rm min}}\right]\,.
\end{align}
This prescription is a semi-implicit realization of radiative drag capable of handling strong RICS recoil. Our RICS module captures radiative drag self-consistently for dipolar field geometries with a single-temperature photon background, while the pair production treatment remains simplified in the 1D geometry (see also Section~\ref{sec:discussion}). In steady state at fixed twist, magnetic energy is converted into particle acceleration, and our RICS model accounts for the subsequent radiative losses of the accelerated leptons.

\section{Results}
\label{sec:results}

We present 1D solutions of plasma dynamics in the quasi-steady state found during the time interval $ct/L\in[5,10]$, following an initial relaxation period $ct/L\in[0,5)$. All simulations use thermal background photons with $k_{\rm B}T=1\,{\rm keV}$.
The main results of Section~\ref{sec:results} are obtained with a mass ratio of $m_i/m_e=100$, and Section~\ref{sec:mass_ratio_multiplicity} examines the dependence of the equilibrium multiplicity on the mass ratio. We discuss limitations of our 1D setup in Section~\ref{sec:discussion}.

\subsection{Phase-Space Structure and Gap Formation Along Magnetospheric Field Lines}
\label{sec:phase_structure_gaps}

\begin{figure*}
 \includegraphics[width=0.49\linewidth]{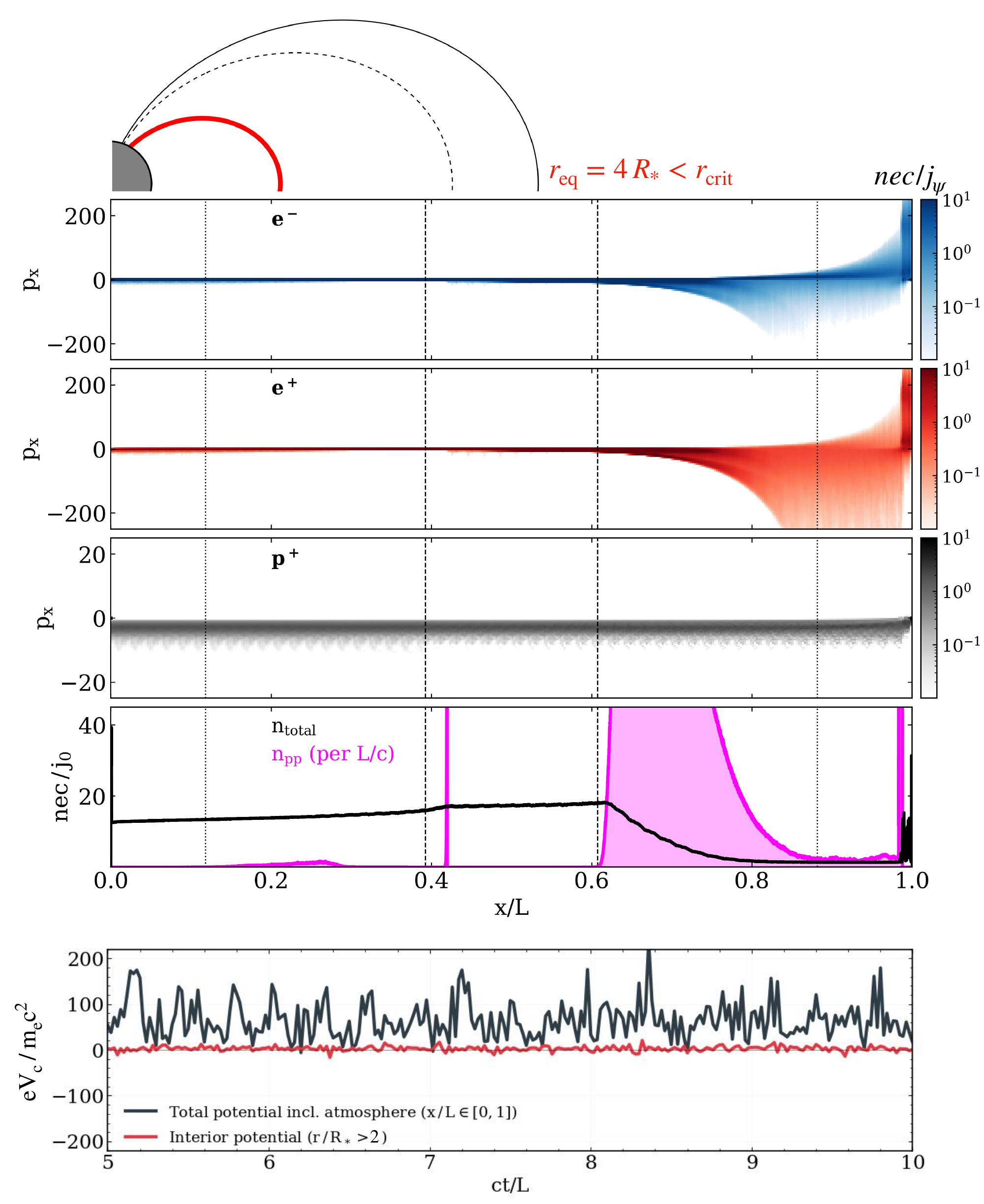}
 \includegraphics[width=0.49\linewidth]{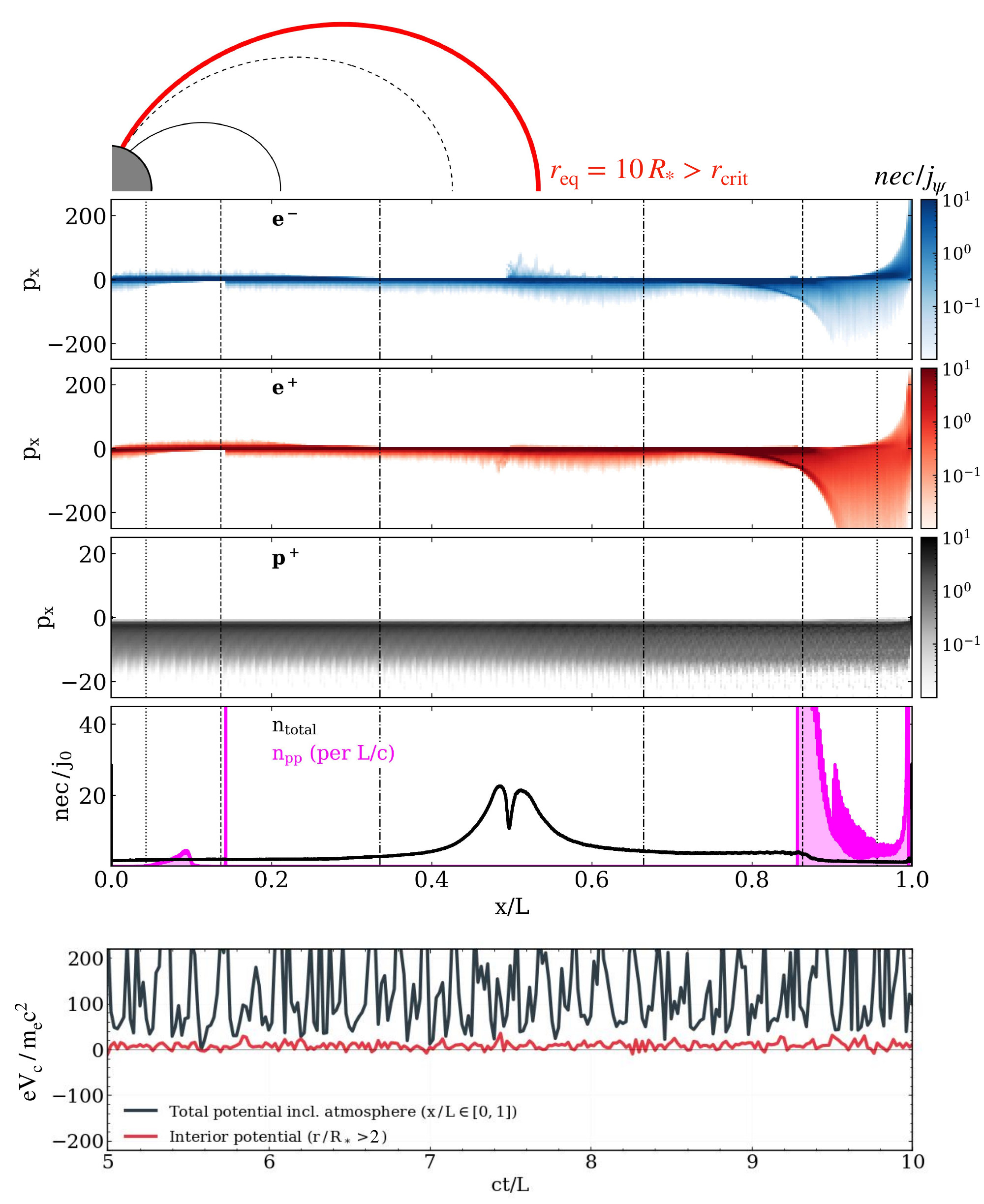}
 \caption{Phase space, multiplicity, and circuit potentials for selected field lines with $\gamma_{\rm min}>1$ (left panels) and $\gamma_{\rm min}\approx 1$ (right panels). We show phase space averages for electrons ($e^-$, first row), positrons ($e^+$, second row), and ions ($p^+$, third row) after initialization, once the plasma reaches a quasi-equilibrium during $ct/L\in[5,10]$. The fourth row shows the average total density of all simulated particles (black line) and the pair production rate averaged over one light crossing time (magenta). The bottom row shows circuit potentials across the full domain (black line), with the red line indicating the same quantity restricted to radii $r>2R_\ast$ on both sides, excluding the near-surface atmospheric region. No large-scale potentials develop in the outer corona. Top panels show magnetic isosurfaces $b=[1,\,0.1,\,0.01]$.}
 \label{fig:FIGURE2}
\end{figure*}

\begin{figure*}
 \includegraphics[width=0.99\linewidth]{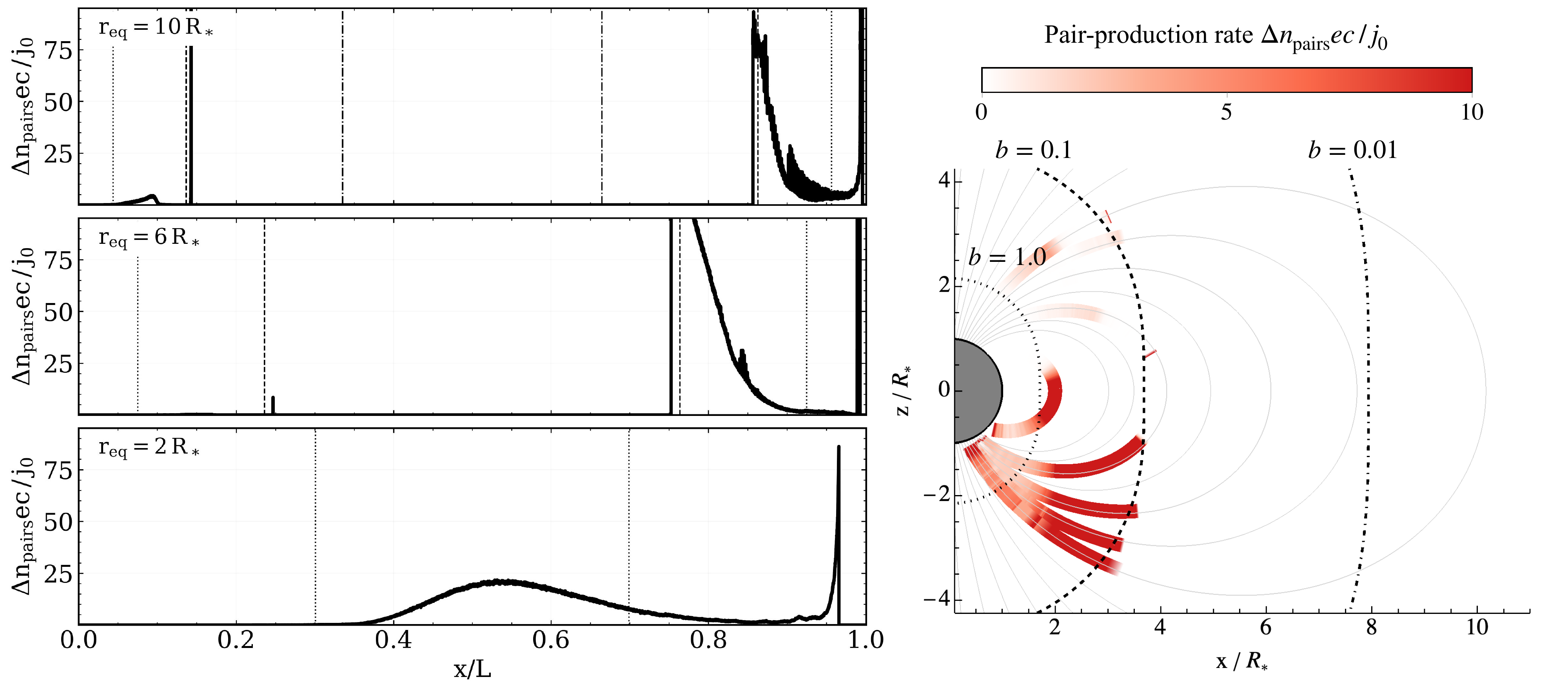}
 \caption{Distribution of pair production locations throughout the magnetosphere. Left: Pair production rate averaged over one light-crossing time in the equilibrium state $ct/L\in[5,10]$ for field lines extending (top to bottom) to $10$, $6$, and $2\,R_\star$. Right: Pair production rate and location in a 2D representation, with dipolar field lines shown as solid grey lines. The color scale is saturated to better highlight the distribution. All panels show magnetic isosurfaces corresponding to $b=[1,\,0.1,\,0.01]$.}
 \label{fig:FIGURE3}
\end{figure*}

To sustain the twist-induced current $j_\psi$, charge carriers are extracted from the atmosphere layer at the domain boundary (i.e., from the stellar surface), with the species and direction of extraction determined by the sign of $j_\psi$. In our simulations, the northern hemisphere provides atmospheric electrons propagating southward (left to right), while the southern hemisphere provides atmospheric ions propagating northward (right to left). We define electric gaps as the localized regions with unscreened fields $E_\parallel$ that strongly accelerate particles. Hard X-rays and secondary $e^\pm$ can be produced outside the gaps, especially for plasma flowing toward larger radii. The drag force tends to reduce the Lorentz factors $\gamma$ of the outflowing leptons toward the attractor $\bar\gamma$ (Equation~\ref{eq:D}). 
Leptons flowing along field lines with $r_{\rm eq}<r_{\rm crit}$ cross the equator with
$\gamma_{\rm min}>1$, and $\gamma>\bar\gamma$
(Figure~\ref{fig:FIGURE1}, left panel, see also Appendix~\ref{app:rics_cross_section}). For
field lines with
$r_{\rm eq}>r_{\rm crit}$, strong RICS drag efficiently slows down leptons to $\gamma_{\rm min}\approx 1$ (Figure~\ref{fig:FIGURE1}, right panel). 

Figure~\ref{fig:FIGURE2} shows the time-averaged plasma parameters measured on two chosen field lines and the history of the total potential drop $V_c$. For the field line with $r_{\rm eq}<r_{\rm crit}$ (left panels), leptons are accelerated mainly in the southern hemisphere, where strong electric potentials are induced to extract ions from the stellar surface. Efficiently accelerated leptons moving towards the star undergo RICS reactions and pair production close to the atmosphere. Leptons moving away from the star encounter RICS drag in the extended magnetosphere, and pair production continues until the local magnetic field drops below $b\approx 0.1$. The pair-loaded outflow then proceeds toward the apex of the magnetic field line at the magnetic equator, crosses the equatorial plane and continues into the northern hemisphere. Once the flow passes the equatorial region with the strongest RICS drag, it moves toward the star without further deceleration. As a result, the pair-loaded flow from the southern hemisphere screens $E_\parallel$ and prevents gap formation in the northern hemisphere. A single gap in the southern hemisphere is sufficient to sustain $j_\psi$ along the entire field line. Remarkably, the gap is localized near the star, right above the atmospheric layer.
There is no notable magnetospheric potential away from the atmosphere (left bottom panel). 

The gap structure is different on
field lines with $r_{\rm eq}>r_{\rm crit}$ (Figure~\ref{fig:FIGURE2}, right panels). Leptons are still efficiently accelerated where atmospheric ions are extracted from the star, and pair production still extends to larger radii until $b\approx 0.1$. However, the drag in the equatorial region is so strong that $\gamma_{\rm min}\approx 1$. Any lepton outflow approaching the equator cannot stream freely and cross into the other hemisphere.
We find that the plasma circuit can sustain $j_\psi$ in two ways. First, for realistic RICS drag times $\eta_{\rm min}=10^3\times(n_\psi/n_0)^{1/2}$, efficient gaps close to the star in the southern hemisphere can accelerate ions that move through the equator at larger velocities without experiencing RICS drag (see Figure~\ref{fig:FIGURE2}, right panel). Fast-moving ions carry velocity and density fluctuations northward, and accelerate leptons in the direction of ion motion via an ion-pair streaming instability. Some leptons cross the strong-drag equatorial plane and screen $E_\parallel$ in the northern hemisphere. This results in an asymmetric configuration with one prominent gap and pair production in the southern hemisphere. Second, for shorter RICS drag times $\eta_{\rm min}=10^2\times(n_\psi/n_0)^{1/2}$, the radiative drag is so strong that no leptons are carried across the equatorial plane. Pairs are instead generated by pair production in both hemispheres 
(Appendix~\ref{app:qedprocesses}, Figure~\ref{fig:FIGURE8}). The resulting gap structure is more symmetric, and plasma flows in both hemispheres toward the equator, where it accumulates and annihilates. In this regime, ions are accelerated less efficiently.\footnote{For $m_i/m_e=1$, the gap structure becomes symmetric between the northern and southern hemispheres. In this case, without energetic ions to mediate lepton transport through the strong-drag region at the equator, the current is sustained by gaps in both hemispheres, and pairs are produced symmetrically regardless of the drag strength.}

In either case, the steady state electric potential remains small in the extended magnetosphere but can be substantially higher close to the stellar surface. As shown in the bottom panels of Figure~\ref{fig:FIGURE2}, the near-surface region dominates the magnetospheric potential (black line), reaching $eV_{\rm c}/m_ec^2\approx 200$ ($ V_{\rm c}\approx100\,\text{MeV}$) and accelerating leptons to energies sufficient for pair production via RICS. The near-surface voltage sets an upper limit on the dissipation rate of magnetospheric twist.

Figure~\ref{fig:FIGURE3} displays pair production rates obtained for individual field lines in the model with $\eta_{\rm min}=10^3\times(n_\psi/n_0)^{1/2}$. While electric gaps are localized near the star, the pair production zones extend to a larger radius where $b\approx 0.1$. One can also see the asymmetry between the northern and southern hemispheres: pair production is far weaker in the northern hemisphere (and
appears only for field lines with $r_{\rm eq}>r_{\rm crit}$). The pair production rate and resulting (a)symmetry is controlled by the ratio of the drag timescale $\tau_{\rm min}$ and the plasma timescale $\omega_{\rm p}^{-1}$ (cf. Appendix~\ref{app:qedprocesses}, Figure~\ref{fig:FIGURE8}).

In some cases, sharp bumps appear in the pair creation rate near the pair production magnetic field threshold (Figure~\ref{fig:FIGURE2}). We suspect these are numerical artifacts caused by the matching between full RICS scattering and the continuous drag force. These features do not substantially affect the pair multiplicity, so we neglect their impact here, but future simulations should revisit this boundary matching with greater care.


\begin{figure*}
 \includegraphics[width=0.99\linewidth]{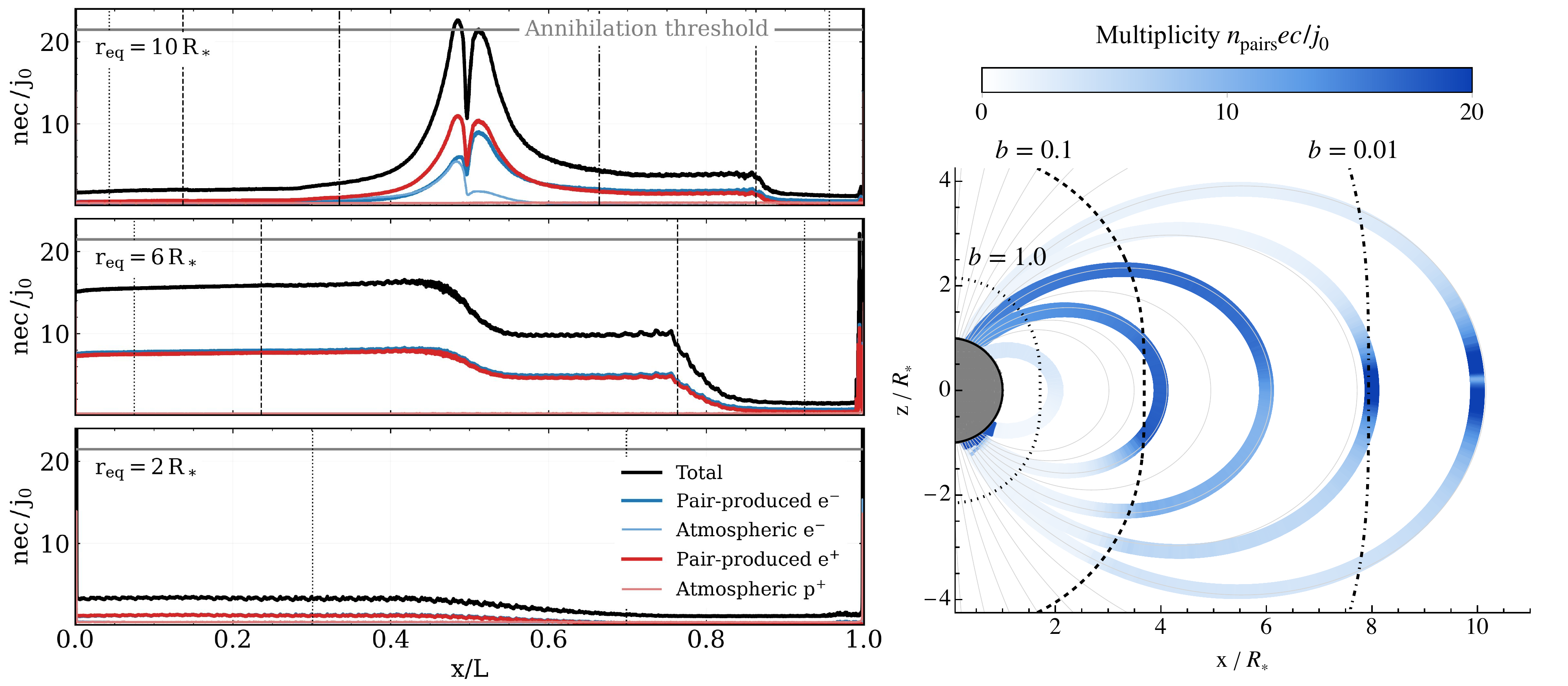}
 \caption{As Figure~\ref{fig:FIGURE3}, but showing the magnetospheric multiplicity distribution. Left: Total particle density averaged over $ct/L\in[5,10]$. Right: Multiplicity and its spatial distribution in a 2D representation.}
 \label{fig:FIGURE4}
\end{figure*}

\subsection{Pair Multiplicity and Mass-Ratio Dependence}
\label{sec:mass_ratio_multiplicity}

Our simulations demonstrate that pair multiplicity is distributed asymmetrically throughout the magnetosphere. Extended RICS zones produce a high-multiplicity pair outflow from a single gap. In the equatorial region, the lepton density increases to maintain current continuity as particles approach their minimum velocity. For field lines with $r_{\rm eq}<r_{\rm crit}$, where the plasma does not slow down completely, a high-multiplicity flow propagates northward toward the star, forming an extended region of pair-dominated plasma. Along field lines with $r_{\rm eq}>r_{\rm crit}$, RICS slows the leptons to a nearly complete stop at the equator, forming a high-multiplicity region symmetric about the equator. There, the highest densities approach the annihilation threshold imposed by the numerical constraints (Figure~\ref{fig:FIGURE4}, top left panel). In our models, this threshold is reached only in localized regions of the highest density. While it may induce transient feedback, it does not significantly affect the time-averaged pair production or gap structure. 

In the equilibrium state, pair-produced particles carry most of the current $j_\psi$. The left panels of Figure~\ref{fig:FIGURE4} show that neither atmospheric leptons nor ions contribute significantly to the total density. Due to the complete slowdown of the plasma at the equator, the $r_{\rm eq}=10R_\ast$ field line retains a subdominant fraction of atmospheric electrons (top left panel of Figure~\ref{fig:FIGURE4}). We conclude that the simulated circuits \emph{ignite} and self-sustain, with atmospheric electrons playing at most a minor role as seeds for pair production. The extracted atmospheric ions play a more significant role by helping current circulation through the region of strongest drag.

\begin{figure}
 \includegraphics[width=0.99\linewidth]{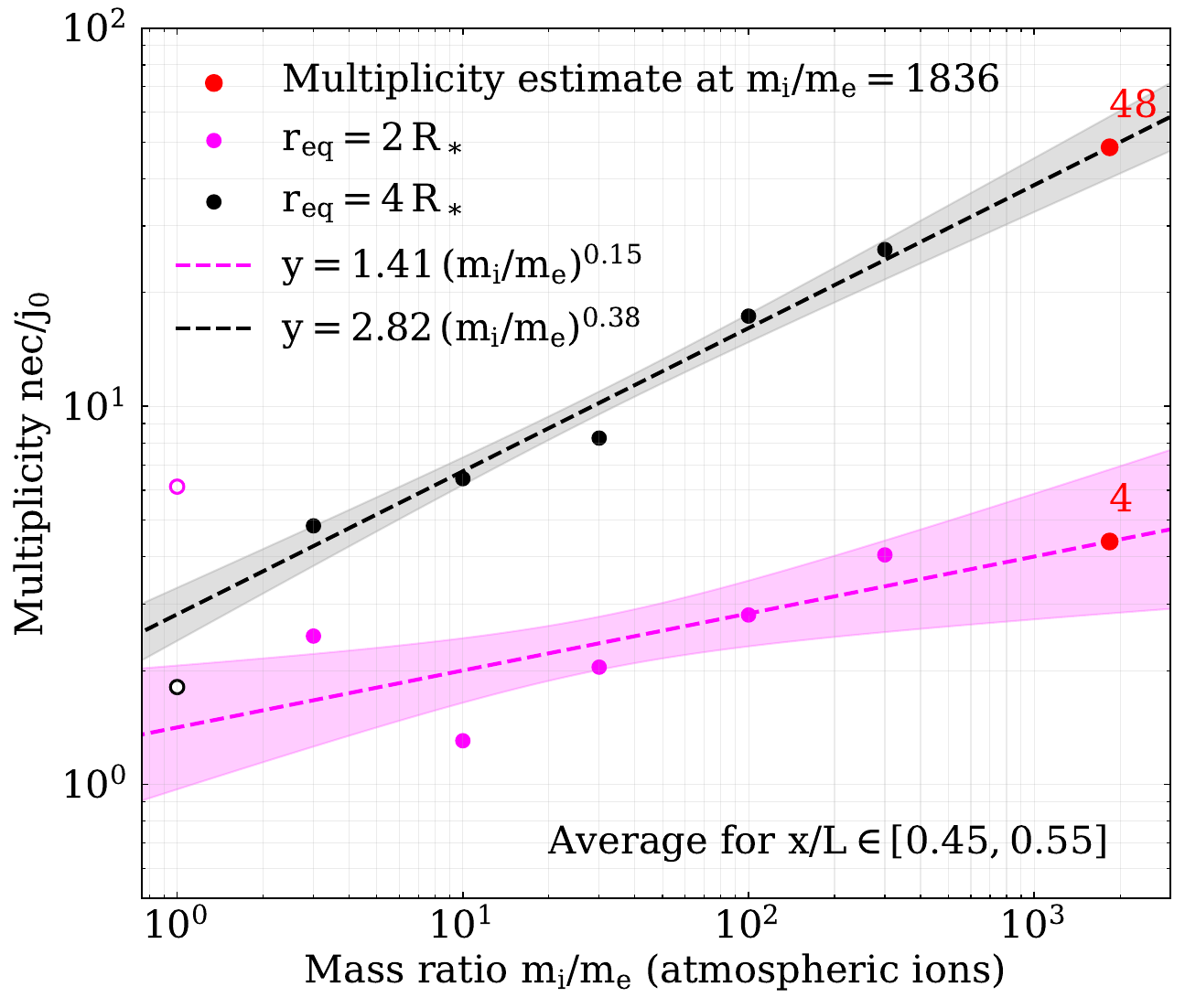}
 \caption{Scaling of the average multiplicity in the equatorial region for field lines with $r_{\rm eq}<r_{\rm crit}$ as a function of ion mass ratio. Data points from PIC simulations are shown as dots, and fitted power laws are indicated by the dashed lines. The extrapolated value at $m_i/m_e=1836$ is shown in red color. Results for $m_i/m_e=1$ are displayed as circles and not included in the curve fitting. The shaded bands denote the $1\sigma$ uncertainty of the power-law fit, obtained via the covariance of the linear regression in log–log space.}
 \label{fig:FIGURE5}
\end{figure}

The maximum acceleration of leptons close to the star controls the number of pairs produced by the RICS cascade. \citet{Beloborodov2013} estimate that the pair cascade produces multiplicities $\mathcal{M}\sim V_\psi\propto (m_i/m_e)^{1/2}$ (Equation~\ref{eq:gamma_circuit}). In PIC simulations with artificial scale separation, the available acceleration increases with the mass ratio $m_i/m_e$. Thus, simulated pair loading and multiplicity distributions approximate the magnetospheric equilibrium, but require rescaling to realistic conditions. For this work, it was computationally prohibitive to model higher mass ratios in the strong drag limit for $r_{\rm eq}>r_{\rm crit}$. Then, the high-density flow in the equatorial region would approach the annihilation threshold, requiring resolutions beyond the currently employed $N_x=3.84\times 10^5$ cells (for $r_{\rm eq}/R_\ast=10$).

Figure~\ref{fig:FIGURE5} evaluates the scaling of maximum $\mathcal{M}$ with mass ratio for field lines with $r_{\rm eq}<r_{\rm crit}$, for which no additional density buildup due to complete plasma slow-down occurs. We vary the mass ratio of atmospheric ions between $m_i/m_e=1-300$ while keeping all other parameters constant. All runs use the same setup as outlined above, except at the highest ion-mass-ratio ($m_i/m_e=300$), where we employ twice the resolution and use a shorter steady-state interval to determine $\mathcal{M}$. Extrapolating to realistic $m_i/m_e = 1836$, the resulting multiplicities can reach $\mathcal{M} \sim 10^2\!-\!10^3$ in the outer corona \citep[as predicted by][]{Beloborodov2013}; the inner corona remains in a low-$\mathcal{M}$ regime.

\section{Discussion and Conclusion}
\label{sec:discussion}

The presented PIC simulations demonstrate how
self-sustained plasma circuits enable long-lived magnetic equilibria in RICS-regulated magnetar magnetospheres. We identify an asymmetric gap structure that adjusts to sustain the required electric current $j_\psi$ against the self-consistent
radiative drag. The amount of pair production and symmetry of the RICS zones depends on the maximum strength of radiative drag (Equation~\ref{eq:tcool}, see Appendix~\ref{app:qedprocesses}). In all cases, we find that the current is sustained by a modest potential drop along the magnetic field lines (Figure~\ref{fig:FIGURE2}), which enables a long lifetime of the magnetospheric twist. Thus, our simulations with the detailed implementation of RICS confirm the basic picture of \citet[][who used 1D PIC simulations with a simplified prescription for pair creation, based on a fixed activation threshold]{Beloborodov2007} and provide new insights into gap locations on different magnetic field lines.

Forming RICS zones and maintaining a balanced plasma supply critically depend on efficient particle acceleration prior to the onset of RICS. As shown in Equation~(\ref{eq:gamma_circuit}), the available acceleration potential scales with the magnetospheric twist, setting the conditions under which leptons can reach the energies required to initiate pair cascades. Capturing this process in PIC simulations introduces strong numerical requirements on resolving sufficiently small plasma skin-depths and achieving large magnetizations. If these scales are under-resolved, particles enter the RICS regime with energies insufficient for pair production, leading to spurious electric fields that would artificially dissipate the magnetic twist.

Hard X-ray production is expected in the
regions of efficient RICS drag acting on $e^\pm$ outflows from the electric gaps \citep{Beloborodov2013}. The pair production threshold $b \gtrsim 0.1$ (Equation~\ref{eq:blimit}) and the condition for the nearly complete stopping of the outflow in the equatorial plane divide the magnetospheric field lines into four types/regions (see Appendix~\ref{app:ppextent}). Region I: $b>0.1$ is satisfied along the entire field line. In this case, a single extended gap forms along the field line, including its part near the equatorial plane. Region~II: $b<0.1$ at sufficient distance from the star, and $r_{\rm eq}<r_{\rm crit}$. Here, the pair production zone cannot cross the equator. The gap forms in the southern
hemisphere near the star. The plasma outflow from the gap is not stopped in the equatorial plane and continues to the northern hemisphere. There, it screens any significant $E_\parallel$, preventing the formation of a second gap. The asymmetry between the two hemispheres is caused by the extraction of ions, rather than electrons, from the southern (anode) footpoint of the magnetic field line. Region~III: $r_{\rm eq}>r_{\rm crit}$. Here, we still find gaps near the star in the southern hemisphere. Although, the drag is extremely strong and practically stops the leptons in the equatorial plane, ions streaming northward carry a small fraction of leptons across the equatorial plane via streaming instabilities and thereby screen $E_\parallel$ in the northern hemisphere. Region~IV: field lines extend very far from the star ($r_{\rm eq}>30 R_\star$) where drag becomes inefficient \citep{Beloborodov2013_eflows}. We did not simulate Region~IV in the present paper; its gap structure is expected to be similar to that in Region~II.

We emphasize the importance of the 
parameter $\eta$ for the magnetospheric circuit, in particular in Region~III where ion streaming instabilities (operating on a timescale $\propto \omega_{\rm p}^{-1}$) play a crucial role. We presented the results obtained with a more realistic large $\eta_{\rm min}$ (Section~\ref{sec:results}); however, we also performed simulations with a smaller $\eta_{\rm min}\approx10^2$ (Appendix~\ref{app:qedprocesses}). The parameter $\eta$
(Equation~\ref{eq:tcool}) is sensitive to two factors: (i) the local plasma temperature (which enters in the third power), and (ii) the local plasma density. We used a conservatively low estimate of density; in reality, densities in the annihilation zone will be considerably higher, giving a shorter skin depth and increasing $\eta$. We therefore consider the models presented in Section~\ref{sec:results} to be the more realistic case. The low-$\eta_{\rm min}$ runs presented in Appendix~\ref{app:qedprocesses} show intermittent gap activity at the equator, competing with gaps closer to the star. This results in a higher voltage in the circuit (and hence a shorter lifetime of the magnetospheric twist). The development of equatorial voltage drops may indicate that the imposed RICS drag in the low-$\eta_{\rm min}$ models is unrealistically strong, and that the simulation may not sufficiently separate the relevant scales.

The locations of gaps determine the directions of $e^\pm$ flows that emit hard X-rays. Photons generated by RICS in Region~I are too energetic for direct escape; they will become reprocessed by the outward pair cascade until it reaches radii where $b\lesssim 0.1$. The emission powered by gaps in Region~I can cover a large part of the magnetosphere. However, Region~I may be completely deactivated, since the lifetime of electric currents in this region is shortest \citep{Beloborodov2009}. In Region~II, the energy of RICS photons scales as $E_\gamma\sim5\times10^{4}\,b^{2}\,\mathrm{keV}$. It drops below $\sim 1$\,MeV at sufficient distance from the star, where $b\lesssim0.1$ allows direct escape without further pair creation \citep{Beloborodov2013}. The fact that the gap is present only in one hemisphere implies an asymmetry in X-ray production. As a result, X-rays from Region~II will be invisible for a large set of unfavorable lines of sights. The predicted phase-resolved spectra of magnetars can be calculated and compared with data using fitting methods similar to \citet{Hascooet2014}. We leave this investigation to future work.

The 1D PIC simulations used in this paper and in \cite{Beloborodov2007} have limitations. The current $j_\psi$ is prescribed and fixed, and $E_\parallel$ is determined by the 1D Gauss law integrated along a magnetic field line. For real magnetars, the transverse structure of the electric circuit becomes important \citep{Chen2017}. In particular, the corrected $E_\parallel$ is reduced compared with the 1D model. We note, however, that the accelerating potentials found in our models are concentrated near the star, where the corrections to $E_\parallel$ are moderate. Therefore, we expect that our conclusions will hold in 2D (axisymmetric) or full 3D models. Our unpublished preliminary 2D models indicate that such simulations require large computational resources because of the demanding resolution constraints.

For the magnetospheric circuit in Region~III, plasma accumulates near the equatorial plane, balanced by $e^\pm$ annihilation. Our implementation of annihilation imposes a maximum plasma density that allows sufficient resolution of the plasma skin depth. The true annihilation rate is lower and will operate at higher plasma densities. Careful implementation of the annihilation regime will be essential also in 2D/3D simulations.

Future work can further improve the implementation of pair production in the magnetar circuit. Our simulations assumed instantaneous pair creation by RICS photons that have sufficient energies for $e^\pm$ creation. This approximation is inherent to the 1D model, where photon emission and pair creation must occur on the same magnetic field line. Actual finite free paths of photons lead to redistribution of pairs in the transverse direction across the magnetospheric field lines, which becomes important near the transition to transparency at $b\sim 0.1$ \citep{Beloborodov2013}. Inclusion of this effect is left to future 2D/3D models. It may also be important to include additional channels of pair creation unrelated to RICS. In particular, collisional QED processes become a strong source of photons if the magnetospheric currents are localized in thin current sheets with high current and plasma densities \citep{Thompson2020,kiuru2026,kiuru2026a}.

Pair discharges in a global 2D/3D circuit depend on the global distribution of the magnetospheric twist, which gradually evolves due to the untwisting process \citep{Beloborodov2009}. RICS further depends on the local intensity of X-rays and their angular distribution. The actual radiation field may differ from the idealized, spherically symmetric radial flow of thermal X-rays assumed in this work. In particular, it is shaped by the neutron star’s anisotropic surface emission--which may include hot spots--as well as radiative transfer through the magnetosphere. A self-consistent treatment in a global model can be achieved using the `virtual beams' method developed by \cite{Beloborodov2013_eflows}.

In summary, RICS-driven pair cascades are a viable self-consistent mechanism for sustaining magnetospheric twists, with a naturally developing asymmetry between the gaps in the northern and southern hemispheres of the magnetosphere.
Our results confirm the main features of the inner/outer corona model of \citet{Beloborodov2011} and extend it by identifying the locations of gaps on different field lines in Regions~I, II, and III. In contrast to the 2D simulations of \citet{Chen2017}, we find gaps located away from the magnetic equator and closer to the star, in particular for field lines forming Regions~II and III that should dominate the observed X-ray luminosity. Pair production zones crossing the equator are found only in the innermost magnetosphere (in Region~I).

\begin{acknowledgements}
We thank Alexander Chernoglazov, Hayk Hakobyan, Amir Levinson, Jazmín Romero-Doldán, Anatoly Spitkovsky, Christopher Thompson, Zorawar Wadiasingh, George Younes, and Owen Young for valuable discussions throughout the course of this work. J.F.M. and A.S. acknowledge support from NSF grant AST-2508744. A.M.B. acknowledges support by NASA grant 80NSSC24K1229, NSF grant AST-2408199, and Simons Foundation grant 446228. L.S. acknowledges support from the DOE Early Career Award DE-SC0023015, NASA ATP 80NSSC24K1826, NSF AST-2307202, and the Simons Foundation (MP-SCMPS-00001470). This work was facilitated by the Multimessenger Plasma Physics Center (NSF PHY-2206609) and supported in part by NSF PHY-2309135 to the Kavli Institute for Theoretical Physics. J.F.M. acknowledges support as a Visiting Professor and Fellow (2025–2027) at the Observatoire de Paris and Université Paris Sciences et Lettres (PSL), hosted by the Laboratoire d’Étude de l’Univers et des Phénomènes Extrêmes (LUX), Sorbonne Université, where this work was finalized. We acknowledge GitHub Copilot for assistance with code development and ChatGPT 5.6 Terra for assistance with figure design and language editing. Simulations were performed on the \textit{Rusty} (Flatiron Institute), and \textit{Discovery} (Dartmouth College) clusters. 
\end{acknowledgements}

\bibliographystyle{aasjournal}
\bibliography{literature.bib}

\appendix

\section{Energetics of Resonant Inverse Compton Scattering}
\label{app:rics_cross_section}

The energetics of resonant inverse Compton scattering (RICS) are discussed in detail by, e.g., \citet{Beloborodov2007} and \citet{Beloborodov2013_eflows}. Here, we outline the main aspects relevant for this work and for the numerical implementation of RICS in PIC simulations. In this section, we denote the lab frame (e.g., the NS frame) by a dash, all other quantities are measured in the excitation and de-excitation frame, respectively.

\subsection{Excitation Phase}
\label{sec:Excitation}

Once a scattering event is triggered, we solve the photon-lepton collision in the lepton rest frame. Energy conservation for a lepton in an elevated Landau level with rest energy $E_B$ reads
\begin{align}
 \gamma E_B=m_e c^2+\hbar \omega'_X\label{eq:exen}\,.
\end{align}
Momentum conservation for the electron momentum $\mathbf{p}_e=\gamma\boldsymbol{\beta}_e$ and a photon momentum with unit direction $\mathbf{\hat{p}}_\gamma$ reads
\begin{align}
\boldsymbol{\beta}_e=\frac{\hbar\omega'_X}{\gamma E_B}\mathbf{\hat{p}}'_\gamma\qquad\Longrightarrow\qquad \gamma =\frac{1}{2}\left[\frac{E_B}{m_e c^2}+\frac{m_e c^2}{E_B}\right]\,.
\label{eq:excitedgamma}
\end{align}

\subsection{De-excitation Phase}

Leptons in the first Landau level can return to the ground state by emitting a photon. In the rest frame of the excited lepton, the photon is emitted at angles $\tilde{\mu}$. According to the two polarization states discussed in \citet[][Appendix~A]{Beloborodov2013_eflows}. With the recoil directed along the magnetic field, the energy of the emitted photon is
\begin{align}
 E_\gamma\left(\theta\right)=\frac{E_B}{\sin^2\theta}\left[1-\left(\cos^2\theta+\frac{m_e^2 c^4}{E_B^2}\sin^2\theta
\right)^{1/2}\right]\,.
\end{align}
The energy of the de-excited lepton is $E_e=E_B-E_\gamma$.
The corresponding lepton momentum is directed in the opposite direction from the photon momentum, and along the magnetic field direction $\mathbf{\hat{b}}=\mathbf{B}/B$, such that $\mathbf{p}_e=-E_\gamma \mu' \mathbf{\hat{b}}$.

\subsection{Energies of Upscattered Photons}
\label{app:energy_upscattered}

For initial (i) lepton velocities $\gamma_i\gg 1$, $\beta_i\approx 1$, we can use velocity addition to write the velocity of the excited (e) lepton in the lab frame using Equation~(\ref{eq:excitedgamma}):
\begin{align}
\begin{split}
 \gamma_e=\gamma\gamma_i\left(1-\beta\right)=\gamma_i \left(\gamma-\frac{\hbar\omega'_X}{E_B}\right)= \frac{1}{2}\gamma_i\left[\frac{E_B}{m_e^2 c^4}+\frac{1}{E_B}-\frac{E_B}{m_e^2 c^4}+\frac{1}{E_B}\right]m_e c^2=\frac{\gamma_i}{E_B}m_e c^2\,.
 \label{eq:scatlim1}
\end{split}
\end{align}
For emission perpendicular to the magnetic field (minimum recoil and maximum photon energy), the lab-frame photon frequency is related to the emission frequency in the rest frame of the excited lepton by
\begin{align}
\begin{split}
 E_\gamma&=\gamma_e E'_\gamma=\gamma_e\left(E_B-m_e c^2\right)=\gamma_i\left(1-\frac{m_e c^2}{E_B}\right)m_e c^2\,.
 \label{eq:perpemit}
\end{split}
\end{align}
For emission along the magnetic field and $\gamma_e\gg 1$ we can write
\begin{align}
 E_\gamma\approx \gamma_e\left(\frac{E_B^2-m_e^2 c^4}{E_B}\right)=\gamma_i\left(1-\frac{m_e^2 c^4}{E_B^2}\right)m_e c^2\,.
 \label{eq:paremit}
\end{align}
We use this estimate to derive Equation~(\ref{eq:blimit}). \citet{Beloborodov2013} estimates $\mu\sim-0.5$, and therefore the resonance condition in Equation~(\ref{eq:resgamma}) becomes $\gamma_{\rm res}\approx 10^2b \,(1\,{\rm keV}/E_X)$. Equations~(\ref{eq:perpemit}) and~(\ref{eq:paremit}) can be evaluated for their scaling with $E_B$ when $\gamma_i=\gamma_{\rm res}$ and $E_X\approx 1\,{\rm keV}$:
\begin{align}
E_\gamma \sim
\begin{cases}
\gamma_i\, b\, m_e c^2 \;\approx\; 10^2\, b^2\, m_e c^2\approx 5\times 10^4 b^2\,{\rm keV}, & B \ll B_Q\,, \\
\gamma_i\, m_e c^2 \;\approx\; 10^2\, b\, m_e c^2\approx 5\times 10^4 b\,{\rm keV}, & B \gg B_Q\,.
\end{cases}
\end{align}

\subsection{Extent of the Pair-Producing Region}
\label{app:ppextent}

\begin{figure*}
 \centering
 \includegraphics[width=0.45\linewidth]{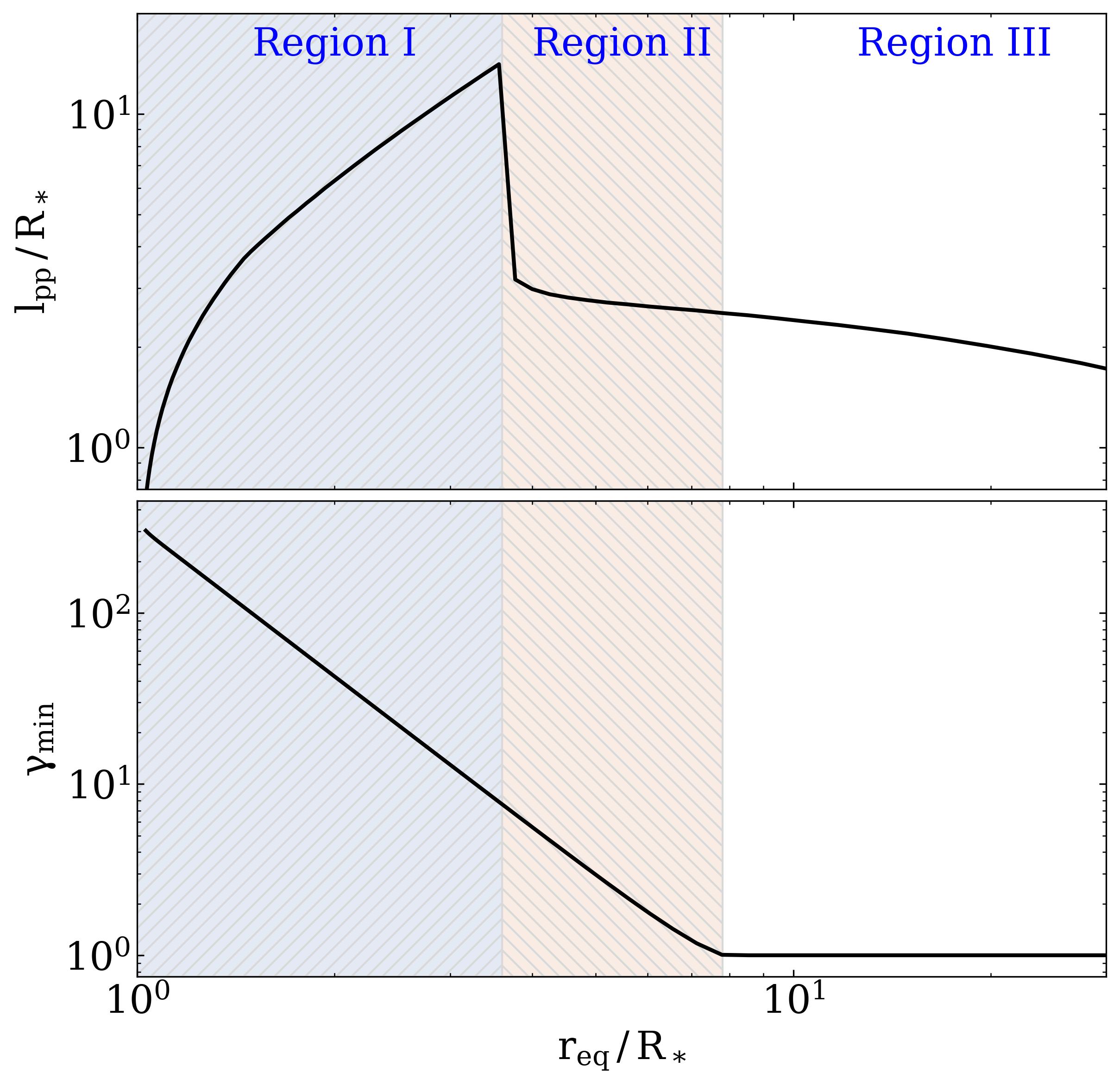}
 \caption{RICS-limited lepton velocities and pair production zone extent in a dipolar magnetosphere with $B_\ast/B_{\rm QED} = 10$ and 1\,keV thermal photons. The top panel shows an estimate of the size of the pair creation zone, defined as the distance between the first scattering of a lepton with $\gamma_0 = 1000$ and the location where the magnetic field drops below $b = 0.1$. For field lines that do not reach regions with $b < 0.1$, the pair production zone spans both hemispheres, leading to a sharp increase in its length (Region I, blue shaded). The bottom panel shows the minimum Lorentz factor $\gamma_{\rm min}$ imposed by RICS drag. The orange shaded region marks a transition regime in which the $b < 0.1$ threshold still limits pair production, while $\gamma_{\rm min}>1$ (Region II). In Region III, pair production extend up to the $b < 0.1$ limit, and the plasma efficiently decelerates to $\gamma_{\rm min} \approx 1$ in the equatorial region.}
 \label{fig:FIGURE6}
\end{figure*}

We use the RICS geometry and associated thresholds to infer the general structure of a dipolar magnetosphere with $B_\ast/B_{\rm QED}=10$, assuming a distribution of $1\,{\rm keV}$ photons emerging radially from the stellar surface (Figure~\ref{fig:FIGURE1}). In a test-particle approach, we integrate particle trajectories along magnetic field lines under the action of the drag force $\mathcal{F}$ defined in Equation~(\ref{eq:contdrag}). By tracking the particle velocity throughout the magnetosphere, we determine the minimum Lorentz factor $\gamma_{\rm min}$ set by RICS drag. For these parameters, the resulting velocity structure is shown in Figure~\ref{fig:FIGURE1} (panel B) and Figure~\ref{fig:FIGURE6}. The strong RICS drag partitions the magnetosphere into two distinct regimes separated by the critical radius $r_{\rm crit}$: field lines along which the minimum Lorentz factor remains above unity ($\gamma_{\rm min} > 1$), and those along which leptons are completely slowed down ($\gamma_{\rm min} \approx 1$).

We also estimate the possible extent $l_{\rm pp}$ of the RICS-mediated pair production region. Assuming leptons are injected with an initial Lorentz factor $\gamma_0 = 1000$, we follow their trajectories along individual field lines to determine the location of the first RICS scattering, marking the onset of the pair cascade. We further identify the point along each field line where the magnetic field drops below the threshold $b = 0.1$. We use the distance between these two locations as an estimate of the size of the pair production region, $l_{\rm pp}/R_\ast$, for a given field line. For field lines that remain entirely within the $b > 0.1$ regime, the pair production region can extend across the equator; in this case, we assume symmetry about the equator to estimate its size. The resulting distribution of $l_{\rm pp}/R_\ast$ throughout the magnetosphere is shown in Figure~\ref{fig:FIGURE6} (top panel). The size increases for short field lines, and then sharply decreases to $l_{\rm pp}/R_\ast \approx 2$.

The pair production threshold $b > 0.1$ (see also Equation~\ref{eq:blimit}), together with the critical radius $r_{\rm crit}$ for complete lepton slow-down, divides the magnetosphere into three distinct regions. \textit{Region I:} the pair production zone can extend across the equator, as $b > 0.1$ is satisfied along the entire field line, and the minimum Lorentz factor remains above unity. \textit{Region II:} the pair production zone cannot cross the equator, and must self-regulate in one hemisphere according to the threshold in Equation~(\ref{eq:blimit}), here taken as $b = 0.1$. In this regime, leptons do not undergo complete slow-down. \textit{Region III:} plasma flows into the equatorial region from both hemispheres, where separate gaps open. Particles experience complete slow-down, reaching $\gamma_{\rm min} \approx 1$ due to strong RICS drag. The distribution and relative extent of these regions depend on the magnetic field strength and the properties of the background photon field. A systematic exploration of this parameter space is beyond the scope of this work and will be addressed in future studies.

\section{Details of the Numerical Setup}
\label{app:setup}

\subsection{Atmospheric Boundary Conditions}
\label{app:atmo_boundary}

We model the stellar surface at radius $R_\ast$ as a perfect conductor with frozen-in magnetic field lines. Surrounding the boundary, pairs and ions are continuously supplied in an atmospheric layer by maintaining the density profile
\begin{align}
 \frac{n_{\rm atmo}}{n_{\rm 0}} = a_0 \exp\left[-\frac{r - R_\ast}{h}\right], \qquad r > R_\ast\,.
 \label{eq:densprofile}
\end{align}
Here $a_0$ denotes the ratio between the maximum atmospheric density and the fiducial density $n_0$, and $h$ is the scale height. The atmosphere is held in hydrostatic equilibrium, where $T \partial_r n_{\rm atmo} = n_{\rm atmo} g$ with a temperature $T$ and a radial gravitational force $g$. This exponential profile corresponds to a constant acceleration $g = -T/h$. In practice, the gravitational force is set to zero once $n_{\rm atmo}/n_0 \ll 1$.

\subsection{Magnetic and Radiation Geometry Along 1D Field Lines}
\label{app:1D_geometry}

\begin{figure*}
\centering
 \includegraphics[width=0.65\linewidth]{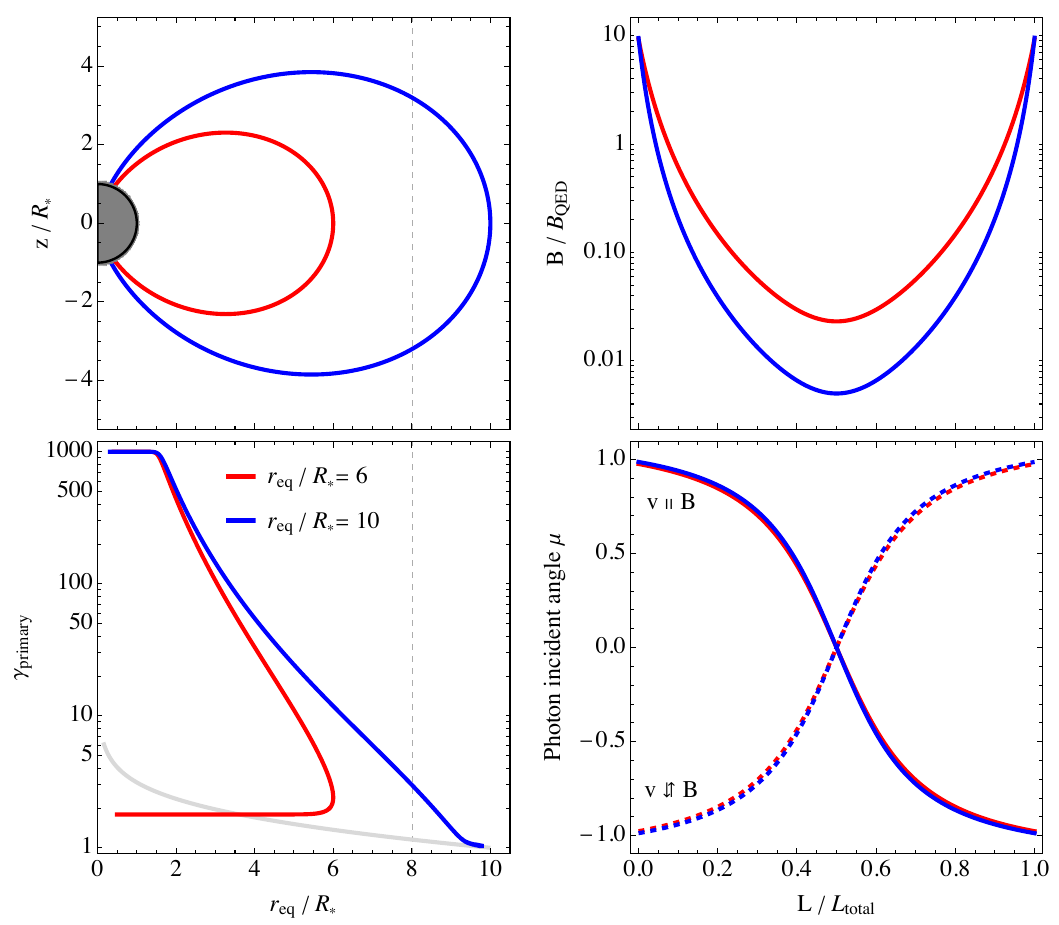}
 \caption{Visualization of the dipolar geometry of magnetic fields and photon properties along isolated field lines. We show the field lines (top left), the velocity evolution of a test particle subject to RICS drag along these lines (bottom left), the magnetic field strength (top right), and the angle between the photon and lepton propagation directions (bottom right). The critical radius $r_{\rm crit}$ is indicated by a dashed line in the left panels. Along dipolar field lines extending beyond this radius, leptons subject to RICS drag decelerate toward the attractor $\mu =\bar{\beta}$ (see Equation~\ref{eq:contdrag}) and eventually come to rest at the equator. The quantities outlined in this figure are provided to the RICS kernel, effectively capturing dipolar geometry within a 1D setup.}
 \label{fig:FIGURE7}
\end{figure*}

In this work, we set up 1D PIC simulations that mimic the global geometry of a dipolar magnetosphere. The radiative reaction follows the dipolar magnetic field geometry and photon distributions expected from a single blackbody of radius $R_\ast$ emitting radially. The dipole magnetic field in spherical coordinates $(r,\theta,\phi)$, for a surface field strength $B_\ast$ is
\begin{align}
 \mathbf{B}_{d}=\frac{B_\ast R_\ast^3}{2}\left(\frac{2\cos\theta}{r^3},\frac{\sin\theta}{r^4},0\right),\qquad B_d=\frac{B_\ast R_\ast^3}{2r^3}\sqrt{3\cos^2\theta+1}\,.
\end{align}
We use field-line-aligned coordinates to parametrize $B_d(l)$ such that $l\in \left[0,L\right]$ tracks the entire length of a field with footpoints located at co-latitude
\begin{align}
 \theta_0= \arcsin\left[\sqrt{R_\ast/r_{\rm eq}}\right]\,.
\end{align}
For the same field line, we derive the angle between surface-emitted photons and the magnetic field, $\mu = B^r / B_d$, and parametrize it as $\mu(l)$. Figure~\ref{fig:FIGURE7} shows these parametrized quantities for the field lines modeled in this work, exemplified by $r_{\rm eq}/R_\ast = 6$ and $r_{\rm eq}/R_\ast = 10$. We also include the effect of the drag force on a single particle entering the magnetosphere with $\gamma_0 = 10^3$ for a blackbody photon temperature of $1\,\mathrm{keV}$.

\section{Dependence on the RICS Drag Time}
\label{app:qedprocesses}

\begin{figure*}
 \includegraphics[width=0.49\linewidth]{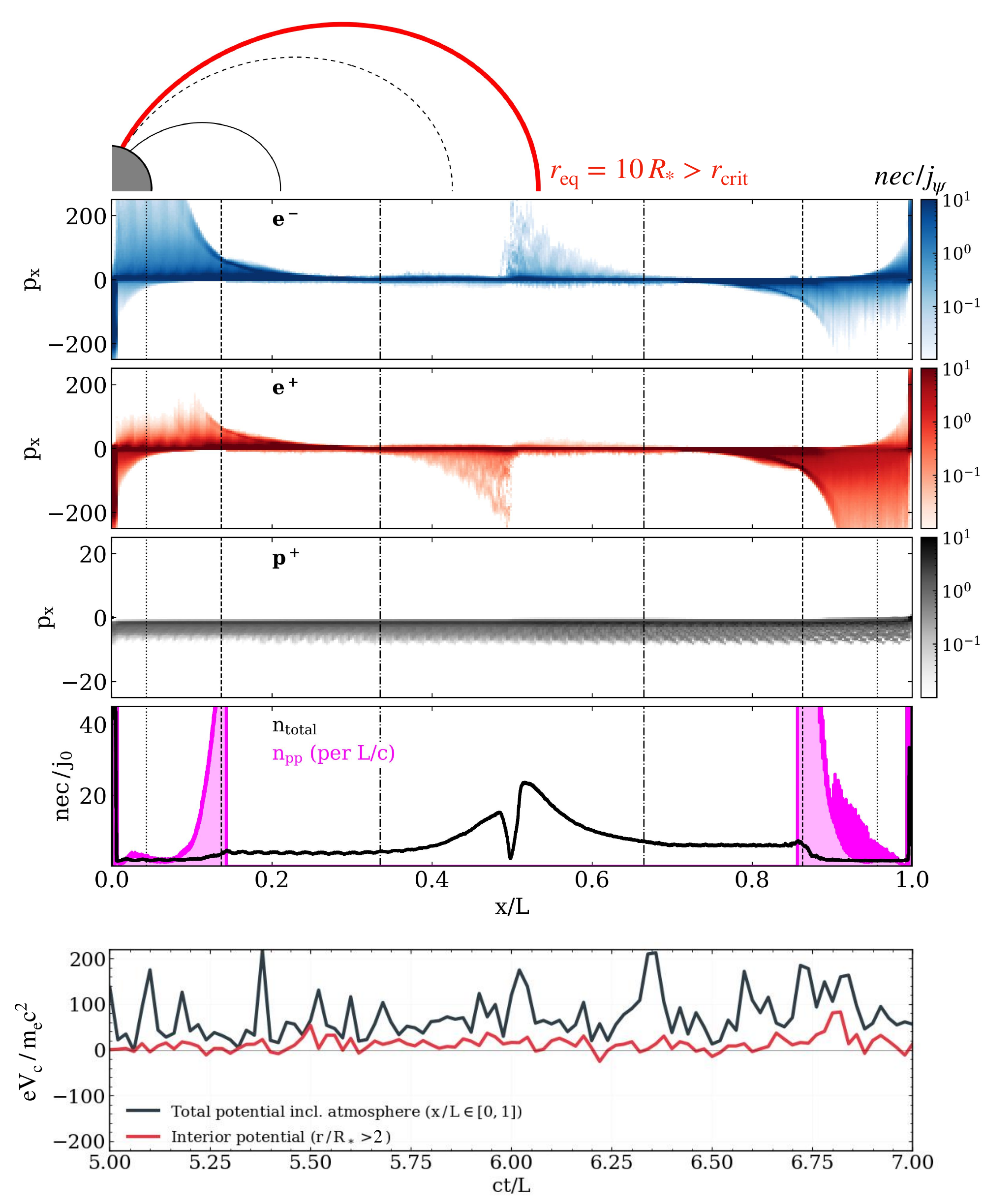}
 \includegraphics[width=0.49\linewidth]{FIGURE2_A2.pdf}
 \caption{Comparison of magnetar circuit realizations for field lines with $\gamma_{\rm min}\approx 1$ for different limits of the fastest RICS drag time. We compare $\eta_{\rm min} = 10^3\times(n_\psi/n_0)^{1/2}$ (right panel, also shown in Figure~\ref{fig:FIGURE2}) with a faster RICS drag timescale, $\eta_{\rm min} = 10^2\times(n_\psi/n_0)^{1/2}$ (left panel). The pair production rate in the southern hemisphere remains mostly unchanged, while for stronger RICS drag, a second gap with substantial pair production opens in the northern hemisphere.}
 \label{fig:FIGURE8}
\end{figure*}

As an additional test, we probe the response of the plasma circuit to shorter RICS drag times (Equation~\ref{eq:tcool}). We use the same parameters as for the cases shown in Section~\ref{sec:simulations} but now limit $\eta_{\rm min} = 10^2\times(n_\psi/n_0)^{1/2}$. Figure~\ref{fig:FIGURE8} directly compares the resulting equilibrium (left panel) with the previously obtained results (right panel, and Figure~\ref{fig:FIGURE2}). 

For shorter RICS drag times, ion velocities and acceleration potentials close to the stellar surface notably decrease. Leptons can no longer be carried to, and in part across, the equator by the northward-moving ions. Instead, an efficient accelerator and pair production gap emerges in the northern hemisphere, resulting in a more symmetric gap structure. Leptons flow into the equatorial region from both sides. The maximum pair multiplicity does not change significantly, though fewer pairs contribute to the current $j_\psi$ in the presence of faster ions (Figure~\ref{fig:FIGURE8}, right panel). Transients with notable $V_{\rm c}$ can occur when the gaps fail to provide sufficient charge carriers to the equatorial region. Then, a fraction of leptons can accelerate at the equator despite strong RICS drag. In our models, numerically imposed pair annihilation likely enhances and in part controls this effect. However, the potentials observed in our simulations do not drive notable long-lived ion acceleration at the equator.

\end{document}